\documentclass[sigconf]{acmart}
\AtBeginDocument{%
 \providecommand\BibTeX{{%
   \normalfont B\kern-0.5em{\scshape i\kern-0.25em b}\kern-0.8em\TeX}}}

\usepackage[utf8]{inputenc}
\usepackage{graphicx}
\usepackage{hyperref}
\usepackage{tabularx}
\usepackage{booktabs}
\usepackage{textcomp}
\usepackage{array}
\usepackage{caption}
\usepackage{multirow}

\newcommand{\figref}[1]{Fig.\,\ref{#1}}
\newcommand{\tabref}[1]{Table\,\ref{#1}}
\newcommand{\secref}[1]{Section\,\ref{#1}}
\newcommand{\eg}{e.g.\ }
\newcommand{\ie}{i.e.\ }

\newcommand*{\model}[1]{\texttt{#1}}

\newcommand*{\corpus}{\texttt{\sc SoMeSci}}
\newcommand*{\corpusl}{Software Mentions in Science}

\newcommand*\rot{\rotatebox{90}}

\renewcommand\footnotetextcopyrightpermission[1]{}
\begin{document}
\fancyhead[OL]{\corpus---\corpusl} 
\fancyhead[OR]{} 
\fancyhead[ER]{David Schindler, Felix Bensmann, Stefan Dietze, Frank Krüger} 
\fancyhead[EL]{}
\title{\corpus{}---A 5 Star Open Data Gold Standard Knowledge Graph of Software Mentions in Scientific Articles
}
%
%
\author{David Schindler}
\email{david.schindler@uni-rostock.de}
\orcid{0000-0003-4203-8851} 
\affiliation{%
  \institution{Institute of Communications Engineering, University of Rostock}
  \city{Rostock}
  \country{Germany}
}

\author{Felix Bensmann}
\email{felix.bensmann@gesis.org}
\affiliation{%
  \institution{GESIS - Leibniz Institute for the Social Sciences}
  \city{Cologne}
  \country{Germany}
}

\author{Stefan Dietze}
\email{stefan.dietze@gesis.org}
\affiliation{%
  \institution{GESIS - Leibniz Institute for the Social Sciences}
  \city{Cologne}
  \country{Germany}
}
\affiliation{%
  \institution{Heinrich-Heine-University}
  \city{Düsseldorf}
  \country{Germany}
}

\author{Frank Krüger}
\email{frank.krueger@uni-rostock.de}
\orcid{0000-0002-7925-3363} 
\affiliation{%
  \institution{Institute of Communications Engineering, University of Rostock}
  \city{Rostock}
  \country{Germany}
}
\begin{abstract}
Knowledge about software used in scientific investigations is important for several reasons, for instance, to enable an understanding of provenance and methods involved in data handling.
However, software is usually not formally cited, but rather mentioned informally within the scholarly description of the investigation, raising the need for automatic information extraction and disambiguation. 
Given the lack of reliable ground truth data, we present \corpus{}---\corpusl{}---a gold standard knowledge graph of software mentions in scientific articles. 
It contains high quality annotations (IRR: $\kappa{=}.82$) of 3756 software mentions in 1367 PubMed Central articles. 
Besides the plain mention of the software, we also provide relation labels for additional information, such as the version, the developer, a URL or citations. 
Moreover, we distinguish between different types, such as application, plugin or programming environment, as well as different types of mentions, such as usage or creation.
To the best of our knowledge, \corpus{} is the most comprehensive corpus about software mentions in scientific articles, providing training samples for Named Entity Recognition, Relation Extraction, Entity Disambiguation, and Entity Linking.
Finally, we sketch potential use cases and provide baseline results.
\end{abstract}


\maketitle              

\section{Introduction}

\begin{figure*}
    {\setlength{\fboxsep}{0pt}
        \framebox{
            \includegraphics[width=\textwidth]{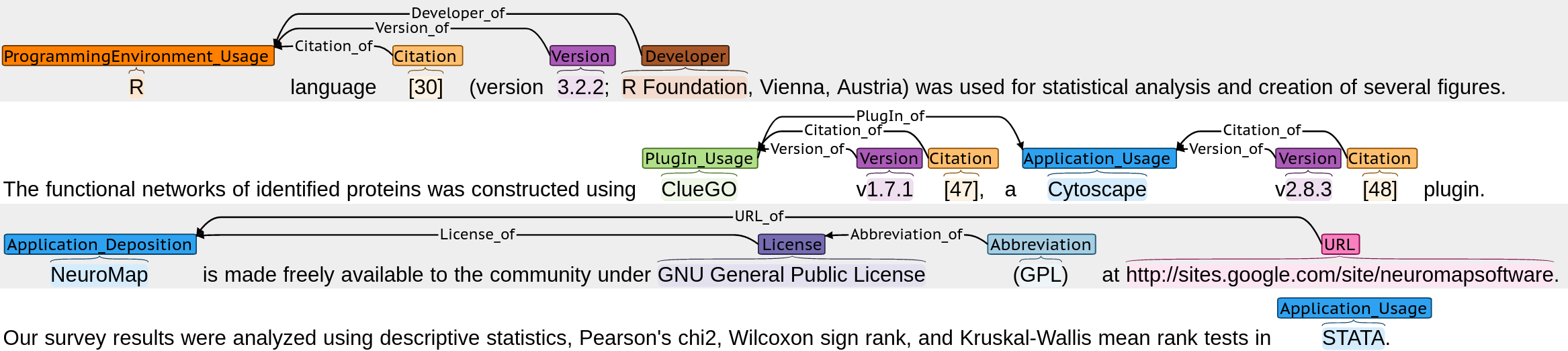}
    }}
    \caption{Example of software mention statements of different completeness and complexity annotated according to the conceptual model described in \secref{sec:conceptual_model}.}
    \label{fig:examples}
\end{figure*}

Software plays a significant role in today's research and is thus often mentioned in scholarly articles.
It is mentioned for different purposes, including its actual usage or the introduction of a novel software, see \figref{fig:examples}.
Knowledge about software used in scientific investigations is important for several reasons, \eg to provide an understanding of provenance of the research results or to give credit to its developers.
Moreover, as software is a key element of data-driven science, knowledge about the particular version, \ie software development state, is a necessary prerequisite for the reproducibility of the scientific results as even minor changes to the software might impact them significantly.
Moreover, information about newly created software is used in publicly funded projects, such as the OpenAIRE Knowledge Graph~\cite{Manghi2019}, aiming at linking Open Access articles to their respective data and software.
Enabling the automatic identification of software creation statements would allow to automatically include such information. 

Despite the existence of software citation principles~\cite{smith2016software,Katz2021}, software mentions in scientific articles are usually informal and often incomplete---information about the developer or the version are often missing entirely, see \figref{fig:examples}.
Spelling variations and mistakes for software names, even common ones~\cite{schindler2020investigating}, increase the complexity of automatic detection and disambiguation.
Training and evaluation of information extraction approaches requires reliable ground truth data of sufficient size, raising the need for manually annotated gold standard corpora of software mentions.
Available corpora~\cite{schindler2020investigating,https://doi.org/10.1002/asi.24454,duck2013bionerds} do not cover all available information, omitting additional information or disambiguation for different spelling variations of the same software.
Moreover, no available corpus distinguishes between the purpose of the mention such as creation or usage (see \figref{fig:examples}).

\begin{figure*}
   \includegraphics[width=\textwidth]{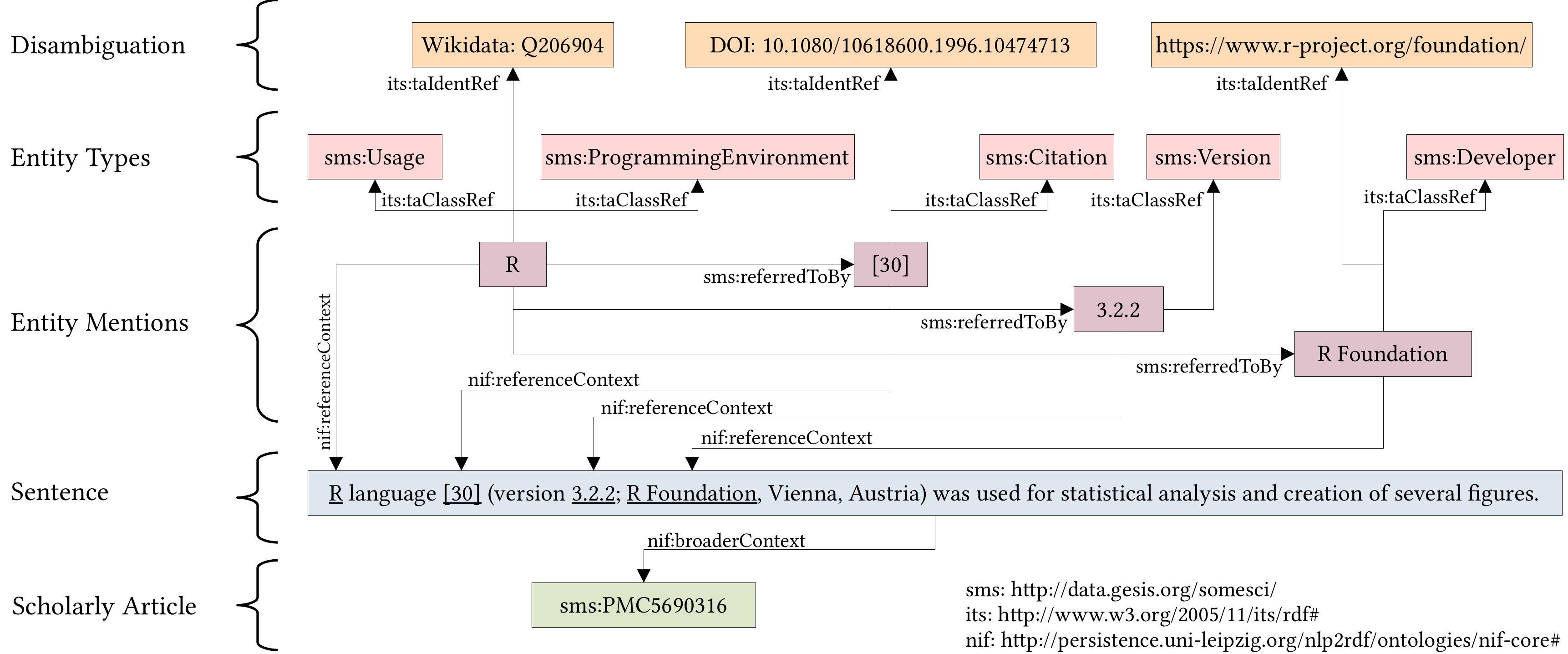}
  \caption{Excerpt of the \corpus{} knowledge graph illustrates the textual references of software mentions and their version, developer and citation. 
  The different levels of representation separate the main concerns of interest, natural language sentences, mentions of entities, their types and disambiguation to knowledge entities.
  For clarity, some information is omitted.}
  \Description{Example software mention statements of different completeness and complexity annotated according to the conceptual model described in \secref{sec:conceptual_model}.}
  \label{fig:teaser}
\end{figure*}

Here, we present \corpus{}, a gold standard knowledge graph of software mentions in scientific articles.
We distinguish between different types of software and mention, provide additional information and disambiguate spelling variations, providing---to the best of our knowledge---the most comprehensive gold standard corpus of software mentions in scientific articles. 
\corpus{} is created by manually annotating 3756 software mentions with additional information, resulting in 7237 labelled entities in 47,524 sentences from 1367 PubMed Central articles.
Data is lifted into a knowledge graph by using established vocabularies such as NLP Interchange Format (NIF)~\cite{Hellmann2013} and schema.org~\cite{ronallo2012html5}, disambiguated and linked to external resources, and shared as a publicly available 5-star Linked Open Data resource as defined by~\citet{5star} that can be explored interactively.
An excerpt of the KG based on the first sentence in \figref{fig:examples} is illustrated in \figref{fig:teaser}.

As shown in Section \ref{sec:usecases}, whereas entity recognition, relation extraction or entity disambiguation with focus on software mentions in scholarly articles require high quality data for training and testing, \corpus{} provides the most comprehensive ground truth dataset so far, able to advance progress in the aforementioned areas. 
We also demonstrate that methods trained with our corpus and their application to unseen corpora of scholarly articles are able to obtain significant insights into software citation practices. 

In the following sections we give an overview of existing corpora for software mentions and introduce a conceptual model for software descriptions in scientific articles and our particular implementation. Then, we describe our annotation process and the \corpus{} corpus that resulted from it. We present reproducible use-cases, discuss the impact, establish baseline results and assess its generalization. Finally, we discuss limitations and conclude. 

\section{Related Work}
\label{sec:relatedwork}
Identification of software mentions in scholarly articles gained momentum over the past years.
Until recently, only few data was available~\cite{Kruger2020}.  
The results of manual content analyses with respect to software in scientific articles had been discussed~\cite{howison2016software,nangia.katz:2017,Allen2018}, and, moreover, only one labeled resource of limited size was available: BioNerDs~\cite{duck2013bionerds} consists of 85 annotated articles and covers mentions of software and databases in the life sciences with a particular focus on bioinformatics.
Only recently, additional resources became available:
SoSciSoCi~\cite{schindler2020investigating} provides 1380 software mentions from 480 methods sections and additional 1005 from individual sentences of 677 of PLoS articles. 
Most recently, Softcite~\cite{https://doi.org/10.1002/asi.24454} was presented with software mentions from 4971 articles split into 2521 biomedicine and 2450 economics articles.
A detailed comparison of the available corpora is provided in \tabref{tab:comparison}.
\begin{table}
    \caption{Comparison of the different available corpora for software mentions in scholarly articles. Domain: BI=Bioinformatics, L=Life Sciences, E=Economics, S=Social Science; IRR: F=FScore, O=Overlap agreement, $\kappa$=Cohen's $\kappa$; {$^\dagger$}=not all negative sentences were published; *=The original paper reported social science as the domain, but adds a remark about possible deviation to the limitations}
    \begin{tabularx}{\columnwidth}{X>{\raggedleft\arraybackslash}p{1.1cm}>{\raggedleft\arraybackslash}p{1.1cm}>{\raggedleft\arraybackslash}p{1.1cm}|>{\raggedleft\arraybackslash}p{1.2cm}}
    \toprule
     & BioNerDs & SoftCite & SoSciSoCi & SoMeSci\\
    \midrule
    \#\, pos.\ Sentences & 1,845 & 3,305 & 1,809 & 2,728 \\
    \#\, neg.\ Sentences & 19,120 & {$^\dagger$}13,069 & 28,045 & 44,796 \\
    \#\,Software & 2,625 &4,093 & 2,385 & 3,756 \\
    \#\,Other &0& 2544 & 0 & 3,481 \\
    \midrule
    Software Type & $\square$ & $\square$ & $\square$ & $\blacksquare$ \\
    Mention Type & $\square$ & $\square$ & $\square$ & $\blacksquare$ \\
    Developer & $\square$ & $\blacksquare$ & $\square$ & $\blacksquare$ \\
    Version & $\square$ & $\blacksquare$ & $\square$ & $\blacksquare$ \\
    URL & $\square$ & $\blacksquare$ & $\square$ & $\blacksquare$ \\
    Citation & $\square$ & $\square$ & $\square$ & $\blacksquare$ \\
    Abbreviation & $\square$ & $\square$ & $\square$ & $\blacksquare$ \\
    Extension & $\square$ & $\square$ & $\square$ & $\blacksquare$ \\
    License & $\square$ & $\square$ & $\square$ & $\blacksquare$ \\
    \midrule
    Disambiguation & $\square$ & $\square$ & $\square$ & $\blacksquare$ \\
    Entity Linking & $\square$ & $\square$ & $\square$ & $\blacksquare$ \\
    \midrule
    Source Domain &BI & L,E & S$^*$ & L\\
    Source Year & 2002--10 & 2000--10& 2007--19 & 2007--20 \\
    IRR & $F{=}.80$&$O{=}.76$&$\kappa{=}.82$& $\kappa{=}.82$, $F{=}.93$ \\
    \bottomrule
    \end{tabularx}
    \label{tab:comparison}
\end{table}

While those corpora have improved the available data for software mentions in scientific articles, there are still aspects not yet covered.
BioNerDs and SoSciSoCi do not take into account information associated with software mentions while Softcite considers \textit{publisher}, \textit{version} and \textit{url}. 
None of the sets distinguish between different types of software and no dataset considers the type of mention. 
Further, no dataset provides any means to disambiguate the software names other than the plain name, which was shown to be insufficient for software disambiguation~\cite{schindler2020investigating}, see also \secref{sec:ie}.

Considering recent paradigm shifts towards data driven analyses across all disciplines making software a first class citizens of science and the resulting changes in awareness about software citations, the publication date of the analysed articles is important to allow conclusions.
While BioNerDs and Softcite articles were both published before 2011, SoSciSoCi contains articles published between 2007 and 2019.

With respect to automatic identification of software mention a similar shift towards supervised learning could be observed~\cite{Kruger2020}.
While Pan et al.~\cite{pan2015assessing} used iterative bootstrapping due to a lack of available data, Duck et al.~\cite{duck2013bionerds} employed manually constructed rules, which were later improved by additional machine learning classifiers~\cite{duck2016survey}.
More recently, machine learning based on Conditional Random Fields (CRFs)~\cite{https://doi.org/10.1002/asi.24454} and Bi-LSTM-CRFs~\cite{schindler2020investigating} has been used.

Knowledge Graphs have been used for the representation of scholarly knowledge to handle the huge amount of information around scientific articles and hidden in the textual descriptions within scientific publications~\cite{10.1145/3360901.3364435}. 
The Microsoft Academic Knowledge Graph (MAKG)~\cite{Faerber2019} covers information about scientific publications and its related meta-data such as authors, institutions, and more.
The Open Research Knowledge Graph~\cite{10.1145/3360901.3364435} represents information about scientific investigations to enable the selection and comparison of those. 
OpenAIRE~\cite{Manghi2019} strives to link scholarly articles with their published data and software.
The automatic construction of such knowledge graphs, however, relies on the existence of automatic information extraction methods trained and evaluated with reliable ground truth data.  

In summary, a high quality annotated corpus of software mentions in recently published, scholarly articles providing labels for 
(1)~types of software,
(2)~types of mention, 
(3)~additional information, as well as
(4)~disambiguation,
is missing but necessary to apply methods of automatic information extraction for automatic, large scale knowledge graph construction.

\section{Data Model}

Software is mentioned within scholarly articles with different levels of detail~\cite{howison2016software}, \eg including version or developer, and for different purposes, \eg usage of an existing application or description of a novel software as illustrated in \figref{fig:examples}.
Different types of software can be distinguished, ranging from end-user applications to plugins or programming environments~\cite{li2017r} (see \figref{fig:examples}), each of them contributing to the investigation in a different way.
The data model was designed to cover these three dimensions by distinguishing between \textbf{Type of Software}, \textbf{Type of Mention}, and \textbf{Additional Information}. 
In the following, both, the conceptual model and the particular implementation are described.

\subsection{Conceptual Model}
\label{sec:conceptual_model}
The model captures the intricacies of in-text software mentions in scholarly articles. 
Definitions for all considered software entities, additional information, and their relationships are provided. 

\subsubsection{Type of Software}

Li et al.~\cite{li2017r} state that beside end-user applications, software packages are commonly used within scientific investigations and propose the categories of \textit{software} and \textit{package}.
Further extending these categories, we distinguish the following software types:

\begin{description}
    \item[\textbf{Application}] is a standalone program, designed for end-users. 
    Using applications usually results in data or project files associated with it, \eg Excel sheets. 
    This definition includes web-based applications such as web-services.  

    \item[\textbf{Plugin}] is an extension to a software that cannot be used individually. 
    Often, the original application can be concluded from the plugin, \eg ggplot2 is a well known R package.

    \item[\textbf{Operating System} (OS)] is a special type of software that is used to manage the hardware of a computer and all other software processes running on it.  
  
    \item[\textbf{Programming Environment} (PE)] is an integrated environment that is built around programming languages and is used to design programs or scripts. 
    This implicitly includes compilers or interpreters.

\end{description}
Distinguishing these software types allows deeper analysis of software usage in scientific investigations, a plugin requires, for instance, the corresponding application.
Software mentions are, therefore, related to one another in two cases: 
(1)~If both, the plugin and the respective application, are mentioned they are considered to be in a \emph{plugin of} relation, and
(2)~if both mentions refer to the same software entity but one provides more additional information it is considered to be a \emph{specification of} the other mention. 

\subsubsection{Type of Mention}
Multiple reasons for software mentions within scholarly articles exist.
Howison and Bullard~\cite{howison2016software} distinguish between \emph{mention} and actual \emph{usage} of a software. 
Based on the further extended categorization scheme of Schindler et al.~\cite{8730730}, we distinguish the following hierarchy of mention types:

\begin{description}
    \item[\textbf{Allusion}] describes each mention of a software name. It does not require an indication of actual usage, \eg stating a fact about the software, or comparing multiple suited software for a problem.
    Software allusions are comparable with scholarly citations used to refer to related work.

    \item[\textbf{Usage} (sub-type of Allusion)] indicates that a software was used, and, therefore, made an actual contribution to the study.
     Software usage statements allow conclusions regarding provenance.

    \item[\textbf{Creation} (sub-type of Allusion)] indicates the introduction of a new software. Software creation statements can be used to conclude the authors of the software and thus to provide scholarly credit to them.
    
    \item[\textbf{Deposition} (sub-type of Creation)] indicates the publication of a new software.
    It extends a software creation statement by stating details about the publication by either providing a license or URL. 
    In the scope of depositions, co-references are used to annotate indirect statements about software when providing details on availability and licensing.  
\end{description}
The mention type can be used to draw conclusions about the role and availability of the software.

Note that an actual software statement within scholarly articles is a combination between a \emph{Type of Software} and \emph{Type of Mention}.

\subsubsection{Additional Information}

According to software citation principles~\cite{smith2016software,Katz2021}, software mentions in scholarly articles should be accompanied by additional information that allows the unique identification of the actually used software.
However, articles often lack this information~\cite{howison2016software}. 
Based on a prior analysis of software mention statements~\cite{schindler2020investigating}, we distinguish between the following additional information about software provided in scholarly articles.

\begin{description}
   \item[\textbf{Developer}] is the person or organization that developed and published the software. 
    Developers can change over time.
 
    \item[\textbf{Version}] indicates a defined state in the software life-cycle. 
    The version is usually provided as a number based identifier.
    As functionality and implementation changes over time, the version is one key element in reproducibility. 
 
    \item[\textbf{URL}] provides a location where additional information can be obtained.
 
    \item[\textbf{Citation}] is formal bibliographic citation connected to the software mention.

    \item[\textbf{Extension}] indicates different function ranges of a software, for instance, distinguishing a professional version from the basic version of a software. 

    \item[\textbf{Release}] is a published version of a software that, in difference to versions, is released using a date based identifier.
    
    \item[\textbf{License}] declares a software's permissions and terms of usage.
    
    \item[\textbf{Abbreviation}] is a shortened form of the software name, often an acronym.
    
    \item[\textbf{Alternative name}] is the opposite of abbreviation and describes a longer name for a given software. Often a software is referred to by its acronym but a more complex, alternative name of the software is provided for completeness. 
\end{description}
The additional information described above is used to specify other entities and is, therefore, always related to another entity. 
Most commonly, they are related to software mentions, but licenses can also be accompanied by a version, a URL or an abbreviation as well as developers who can be specified by a URL or an abbreviation. 

\subsubsection{Entity Identities}
Each textual mention describes a particular software, see for instance \figref{fig:teaser}, providing the actual entity identity.
Beside software, such identities do also exist for developers, licenses and citations.

\subsection{RDF/S Vocabulary}

\begin{figure*}
    \includegraphics[width=\textwidth]{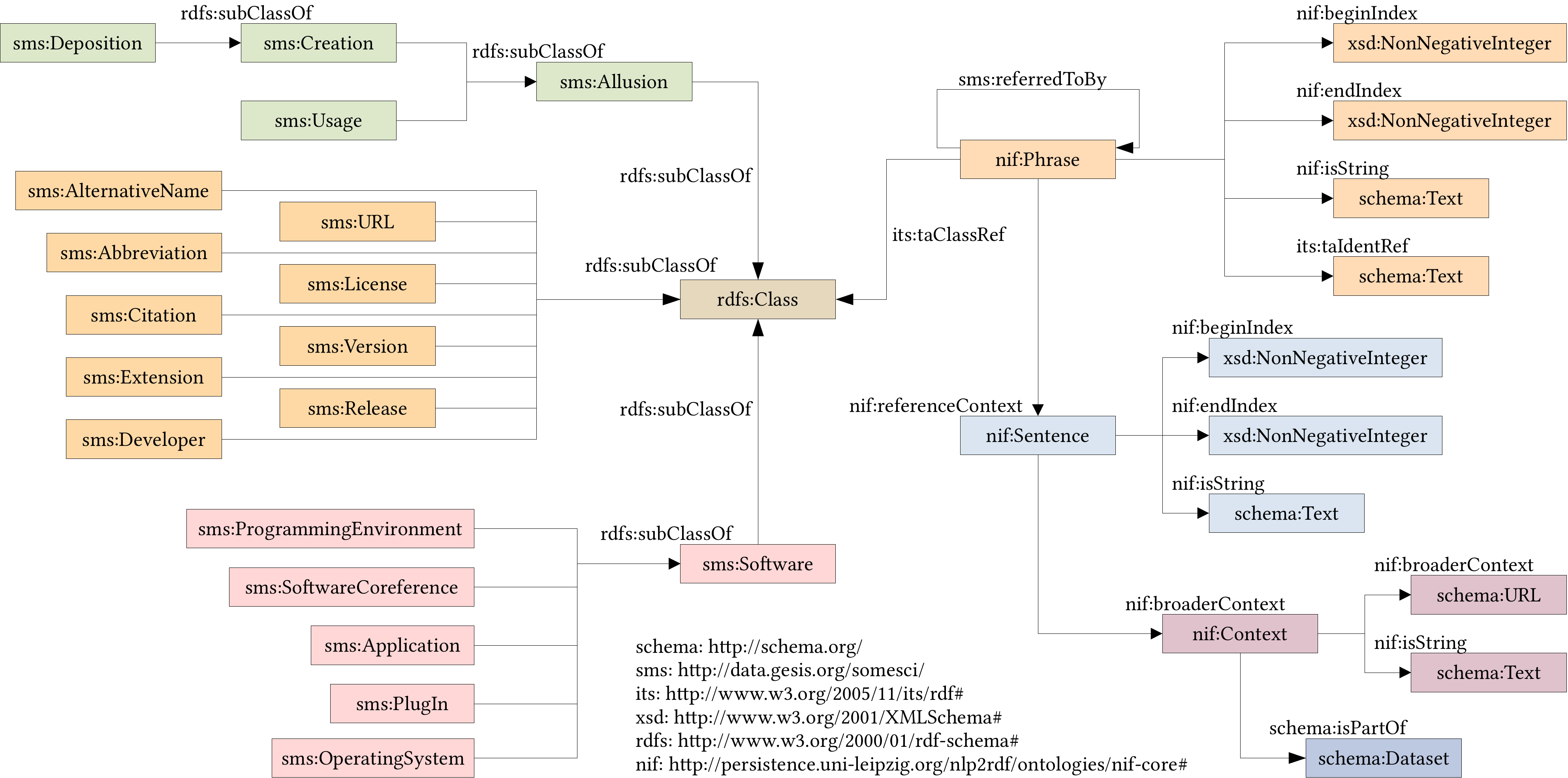}
    \caption{Overview of the data model. Note that properties specialised via \model{rdfs:subPropertyOf} of \model{sms:referredToBy} are used to represent the actual type of relations between entity mentions (\model{nif:Phrase}).}
        \label{fig:datamodel}
\end{figure*}

Objective of the data model is to represent the textual information, including their relations provided from the scholarly articles.
This does not necessarily reflect the truth in case incorrect information is provided by the software mention.
Moreover, as the corpus shall provide a means for training of information extraction and disambiguation methods, modeling was done based on textual entities.
To this end, the data model was based on NIF~\cite{Hellmann2013}, rather than vocabulary for the description of software such as the Software Ontology~\cite{Malone2014}, OKG-Soft~\cite{Garijo2019}, and SoftwareKG~\cite{schindler2020investigating}.
The data model is illustrated in \figref{fig:datamodel}.

In detail, each sentence is represented by type \model{nif:Sentence}, \model{nif:Context} and \model{nif:OffsetBasedString}, providing the properties to embed the entity into a \model{nif:broaderContext} such as the source article. 
The particular position of the respective sentence within the \model{nif:broaderContext} is given by \model{nif:beginIndex} and \model{nif:endIndex}, to allow the reconstruction of the source text (see \secref{sec:article_selection}) to facilitate using the resource for other NLP-based analyses. 
The actual text of the sentence is provided by \model{nif:isString}.

Labelled entities are represented by objects of types \model{nif:Phrase} and \model{nif:OffsetBasedString}, and provide properties for the location of the actual mention within the sentence.
As can be seen from \figref{fig:datamodel}, \model{sms:referredToBy} (\model{rdfs:subPropertyOf} of \model{nif:inter}) reflects the associations between the respective software (or license or developer) mention and the provided additional information.

Class labels, \eg \model{Application} or \model{Usage}, were provided via \model{its:} \model{taClassRef}, where each entity type in \secref{sec:conceptual_model} is represented by an individual \model{rdfs:Class}.
Software type labels represent a combination of software type and mention type, and are thus linked to both.
Furthermore, \model{nif:anchorOf} provides the actual entity mention and \model{nif:referenceContext} links to the sentence with the current mention.
To facilitate entity disambiguation and linking, each entity of classes Software, Developer, License, and Citation provides a link to the corresponding entity with \model{nif:taIdentRef}, as described in \secref{sec:annotation}. 
\figref{fig:teaser} illustrates an example.

The labelled text is lifted to RDF by using rdflib v5.0.0 and serialized into JSON-LD with rdflib-jsonld v0.5.0 by using version 3.7.3 of the Python programming environment.
The descriptive statistics of the dataset is given below.

\section{Labeling Software in Scientific Articles}\label{sec:annotation}

\subsection{Article Selection}
\label{sec:article_selection}
A broad range of scientific publications was selected to establish a corpus of articles.
All selected articles are available from PubMed Central (PMC) Open Access (OA) subset, the largest collection of OA articles.
PMC articles are available in an XML format and were obtained via bulk download on January 22, 2021. 
The main focus of journals indexed in PMC are life sciences, but some, such as PLoS, also publish interdisciplinary articles of other domains.

\begin{figure}
    \centering
    \includegraphics[width=\columnwidth]{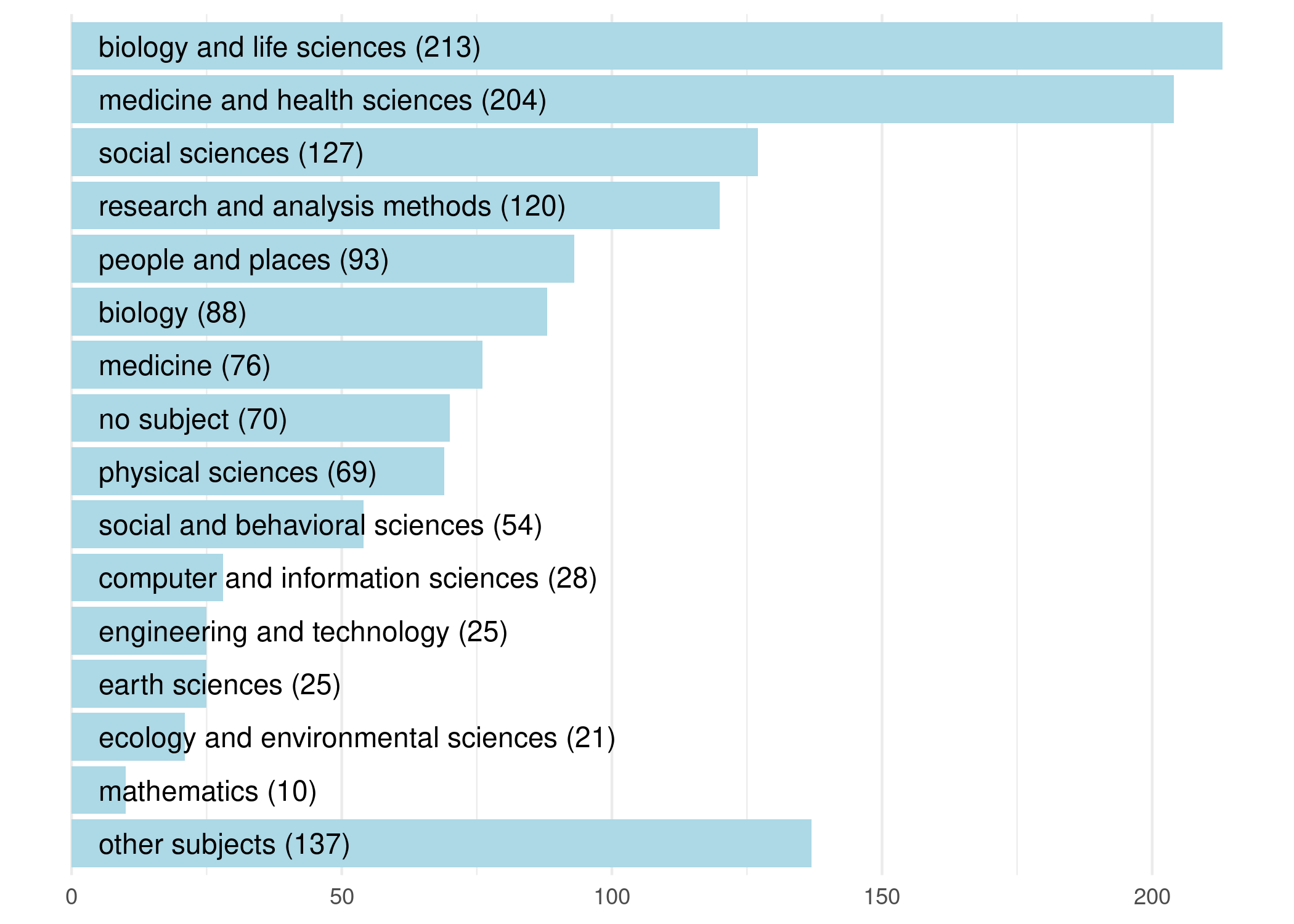}
    \caption{Research subjects of articles in PLoS methods.}
    \label{fig:domain_dist}
\end{figure}

Part of the corpus was created by re-labelling the SoSciSoCi corpus~\cite{schindler2020investigating} that only covers a small aspect of the conceptual model (see \secref{sec:relatedwork} and \secref{sec:conceptual_model}).
Performing a re-annotation of an existing corpus has the benefit of enhancing quality by reducing the number of missed annotations and enforcing consistency between all annotations because articles are read and annotated twice. 
The respective provenance is documented via \model{prov:wasDerivedFrom} in the datamodel.
Overall, the corpus contains 1367 articles split into the following four groups:
\begin{description}
    \item[PLoS methods] 480 methods sections from PLoS journal articles originating from SoSciSoCi.
    Note that while the original study targeted at the social sciences, \figref{fig:domain_dist} illustrates a more diverse distribution of the research domains. 

    \item[PLoS sentences] Individual sentences from 677 PLoS articles selected based on the appearance of common software names originating from SoSciSoCi.

    \item[PubMed fulltext] 100 randomly selected full text articles from PMC OA to allow generalization assessment from methods sections. 
    100 full text articles was considered sufficient as the annotation effort increases with the length of articles.
    
    \item[Creation sentences] Individual sentences from 110 articles that contain software creation statements, selected from PMC OA (n=50) and PLoS (n=60).
    This set was specifically selected to represent the rare but important group of software creation and depositions which are necessary to gain an overall understanding of the software infrastructure in science (see \secref{sec:conceptual_model}). 
    Articles were selected by analyzing titles and abstracts. 
    Matched articles were manually verified and added to the set.
    Individual sentences were then selected by manually review.
\end{description}

\subsection{Annotation Process}
\label{sec:annotation_process}
The conceptual model (see \secref{sec:conceptual_model}) was implemented in the web-based annotation software BRAT v1.3~\cite{stenetorp-etal-2012-brat} that was used to perform all textual annotations, see \figref{fig:examples}. 
The textual annotation process was split into two steps to reduce complexity and minimize the risk for errors during annotation:
(1)~all articles were initially annotated for software type, mention type and relations between two software mentions, and 
(2)~additional information was added for all software annotations resulting from the first step.
For annotation, the articles were split between two of the authors, both having a background in computer science and experience in working with software mentions. 
Additionally, inter-rater reliability (IRR) was assessed by performing overlapping annotations (see \secref{sec:irr}), and resulting differences were discussed and afterwards merged.

After annotation, all \textit{software}, \textit{developers}, \textit{citations} and \textit{licenses} entities were manually disambiguated to unique identifiers on mention basis, partitioned to 4 annotators. 
As sources for identifiers we considered WikiData, package repositories such as CRAN or PyPi, Github, and websites in the stated priority. 
As the initial IRR was found to be insufficient (see \secref{sec:irr}), disambiguation was repeated by 2 annotators, each working on the entire set of mention entities. Results of both were merged ensuring that all links to repositories were adjusted to the suggested canonical form. 

We estimate the overall effort with about 480 hours.
The annotation of PLoS methods sections took up the largest part with 320 hours. 
We estimate an average duration of 20+5 minutes for the two annotation runs for each article followed by 15 minutes of disambiguating 3 software mentions on average, incl. their developers and citations. 
Annotation of PMC fulltexts took 112 hours with an estimated 45 minutes for both annotation passes followed by 22 minutes for disambiguation. 
For annotating individual sentences we estimate an average of 3 minutes per article and for creation sentences 10 minutes totalling in 34 and 18 hours, respectively. 

\subsection{Inter-rater Reliability}
\label{sec:irr}
IRR was determined on 10\% overlap of articles from PLoS methods, PLoS sentences and PubMed fulltexts as described in \secref{sec:article_selection}. 
Creation sections were not considered because the annotations process differed from the other sets as described in \secref{sec:article_selection}.
The IRR was assessed by FScore, as suggested by Hripcsak and Rothschild~\cite{hripcsak2005agreement}, as agreement by chance can be neglected in exact match text annotations, and Cohen's $\kappa$ as categorical errors can still be present within the annotation categories.
Note that partially overlapping annotations were not considered as matching.

Entity annotation for the first step, including both software type and mention type, resulted in a $\kappa{=}.82$.
Detailed results for agreement are given in \tabref{tab:irr_software}.
Operating systems, programming environments and creation statements have perfect agreement.
Agreement is also high for application and usage with 0.94\,FScore and 0.95\,FScore.
For plugin and allusion it was lower but with 0.72\,FScore and 0.84\,FScore still on a reasonable level. 
\begin{table}
\centering
    \caption{IRR of software and mention type. Note that n can differ between both annotations and is provided to give an estimate of the number of annotations. Only matching the most specific sub-type is considered correct for mention type. }\label{tab:irr_software}    
    \begin{tabularx}{\columnwidth}{@{}X*{2}{>{\raggedleft\arraybackslash}p{.75cm}} | X  *{2}{>{\raggedleft\arraybackslash}p{.75cm}}@{}}
        \toprule
        \multicolumn{3}{c|}{Software ($\kappa{=}.82$)} & \multicolumn{3}{c}{Mention ($\kappa{=}.79$)}\\
        \midrule
        Type & F1 &n & Type & F1 & n\\
        \midrule
        Application&0.94&267& Usage & 0.95 & 315\\
        Plugin &0.72&26& Allusion & 0.84 & 10\\
        OS&1.00&8& Creation & 1.00 & 4\\
        PE&1.00&28& Deposition&–&0\\
        \midrule
        Overall & 0.93&329& Overall & 0.95 & 329\\
        \bottomrule
    \end{tabularx}
\end{table}

For the second step, annotation of additional information, such as the version or developer, annotators were provided with the same base annotation of software mentions. 
Objective was to label all additional information and their reference.
In summary, inter-rater reliability for the additional information and relations was almost perfect ($\kappa{=}0.92$ and $\kappa{=}0.94$).
Details are provided in \tabref{tab:additional}.
\begin{table}
\centering
    \caption{IRR for additional information about software. Note that plugin and specification are relations between software entities.}\label{tab:additional}    
    \begin{tabularx}{\columnwidth}{@{}X*{2}{>{\raggedleft\arraybackslash}p{1.2cm}} |  *{2}{>{\raggedleft\arraybackslash}p{1.2cm}}@{}}
        \toprule
        \multicolumn{3}{c|}{Recognition ($\kappa{=}0.92$)} & \multicolumn{2}{c}{Relation ($\kappa{=}0.94$)}\\
        \midrule
        Type& F1 &n & F1 &  n\\
        \midrule
        Abbreviation &1.00&5  & 1.00 & 5\\
        Developer&0.93&85&0.98 & 78\\
        Extension &1.00&4& 1.00 & 4\\
        Alternative name&1.00& 2  & 1.00 & 2\\
        Citation&0.92&40 & 0.98 & 33\\
        Release&0.89& 5 & 1.00 &  4\\
        URL &1.00&16 & 0.88 & 16\\
        Version&0.98& 136 & 0.99 & 132\\
        \midrule
        Overall& 0.96 &293 &0.96 &269\\
        \midrule

        Plugin &–&– &1 & 11\\
        Specification &–&– & 1 & 5\\
        \bottomrule
    \end{tabularx}
\end{table}

The IRR for disambiguation was initially assessed on 10\% of software entity annotations with 68\% agreement.
As described in \secref{sec:annotation_process}, final disambiguation was obtained by merging the result of two annotators. 
The following reasons were found for the low reliability:
(1) different websites for the same resources, 
(2) resource locations changing over time, 
(3) canonical links vs standard links, 
(4) old software without valid, persistent identifier, and 
(5) information behind pay walls.

\section{\corpus{} Knowledge Graph - Facts \& Availability}\label{sec:available}

The \corpus{} knowledge graph consists of 399,942 triples representing 47,524 sentences from 1367 documents from 4 datasets.
2728 of those sentences contain at least one entity, resulting in 7237 labelled entities of the different types. The remaining sentences are included as negative examples.
In total, 3756 software mentions of all types were found. 
A detailed overview of the frequencies of the different label types is provided in \tabref{tab:numbers}.
\begin{table*}
    \caption{Overview of the frequency of the different label types. Coreference refers to an annotated software coreference in the scope of a software deposition statement as described in~\secref{sec:conceptual_model}. Note that only the most specific Mention Type is counted.}
    \label{tab:numbers}
    \begin{tabularx}{\textwidth}{X|*{4}{>{\raggedleft\arraybackslash}p{.6cm}}|*{11}{>{\raggedleft\arraybackslash}p{.58cm}}}
        
        \toprule
        Software Type & \multicolumn{4}{c|}{Mention Type} & \multicolumn{10}{c}{Additional Information}\\
        \midrule
        & \rot{Allusion} & \rot{Usage}  & \rot{Creation} & \rot{Deposition} & \rot{Version} & \rot{Developer} & \rot{Extension} & \rot{Alt.\ Name} & \rot{Citation} & \rot{Release} & \rot{URL} & \rot{License} & \rot{Abbreviation} & \rot{Specification} & \rot{Plugin}\\ 
        \midrule
        Application & 193 & 2206 & 197 & 102& 1221 &766 & 47 & 31 &407 &50 &236 & 34 & 68 & 60 & 68\\
        Plugin & 51 &  275 & 70 & 29&30&12&–&4&141&1&54&12&12&–&6\\
        OS & 30 & 143 &–&–&31&9&4&–&–&1&–&–&–&–&–\\ 
        PE & 47 & 374 & – & – &70&73&–&–&42&35&8 &– &3 & 5 & 134\\
        Coreference &–&–&–&39&–&–&–&–&1&–&40&6&–&–&–\\
        \midrule
       Overall & 321 & 2998 & 267 & 170 & 1352 & 860 & 51& 35 & 591 & 87 & 338 & 52&83&65&208\\
        \bottomrule
    \end{tabularx}
\end{table*}

For each software, as well as for developers, citations, and licenses, the identity was annotated to provide a means for entity disambiguation, see \secref{sec:annotation}.
If available, a URL was used to further provide a means for Entity Linking tasks.
Out of 637 unique software for all types, 432 unique citations, and 9 unique licenses, 22 respectively 3 and 1 could not be linked. 
All 10 unique operating systems and 24 unique programming environments could be linked.

The dataset is available in three formats:
\begin{itemize}
    \item a SPARQL endpoint (\url{https://data.gesis.org/somesci}) with an overview and example queries of the resource to interactively explore the knowledge graph, 
    \item an RDF graph serialized into JSON-LD format for archival to ensure long-time availability of the resource, and 
    \item the raw files that were created by manual annotation and used to create the above graph. 
    The annotation of entities and relations are provided in stand-off format, while the entity's identity and links are given in JSON.
\end{itemize}
The latter two are available for download under CC-BY 4.0 license from Zenodo~\cite{somesci_data}.
All source code used to process the texts, the annotation and the links is publicly available at Github (\url{https://github.com/dave-s477/SoMeSci_Code}).
This also includes a tutorial and code for replicating the use cases described below.
Along with the SPARQL Endpoint we also provide the queries used for all analyses described in \secref{sec:soft_in_schol} as a starting point for further exploration.
We encourage community feedback through Github issues as it provides a transparent way to incorporate resource updates that will consistently be integrated in the KG.

\section{Use Cases and Impact} 
\label{sec:usecases}
As described above, \corpus{} is the first dataset of software mentions in scientific articles that classifies software into respective types, and provides additional information about the software and their identity, paving the way for various use cases, including NLP tasks concerned with scholarly software use or the analyses of software citation practices.
This can either be done by directly analyzing \corpus{}, or through supervised NLP models when used as training/test data.
In the following, we first illustrate the training of information extraction models by providing baseline recognition results and then sketch analyses concerned with the completeness of software citations, facilitated through models trained on our corpus.
Note that our baseline implementation serves to illustrate potential use cases for \corpus{}, rather than establishing state-of-the-art recognition performance.

\subsection{Facilitated Information Extraction Tasks}\label{sec:ie}
As shown in Section \ref{sec:relatedwork}, information extraction from scholarly articles is a crucial challenge requiring targeted models for specific kinds of scholarly information. 
In particular, extraction and disambiguation of software mentions through supervised models has been an area of increasing interest.
Given the comprehensiveness of our model and data, containing marked mentions in natural text, their types and disambiguation to actual software artefacts as defined in a background knowledge base (Wikidata),  \corpus{} facilitates three basic use cases: 
(1)~Named Entity Recognition (NER),
(2)~Relation Extraction (RE), and
(3)~Entity Disambiguation (ED).
In this section, we introduce baseline implementations facilitated by our data. 
For training, the corpus of articles was split into training, development and test set in a $60{:}20{:}20$ ratio for each individual subset of files (see \secref{sec:article_selection}).
The 677 individual PLoS sentence were not split but instead added to the train set to prevent the overestimation of the identification due to the ratio shifted towards common software names.

\subsubsection{Named Entity Recognition}\label{sec:NER} NER is performed for:
(1)~Software in general, 
(2)~software type, 
(3)~mention type, and 
(4)~all additional information.
The baseline is established by the Bi-LSTM-CRF model described in Schindler et al.~\cite{schindler2020investigating} and a separate model is trained for each task.
Articles are tagged sentence-wise and each sentence is processed as a sequence of tokens.
Tokens are transformed into feature representations by concatenating word embeddings~\cite{mikolov2013distributed} trained on PMC OA and character based feature representations calculated by a Bi-LSTM to learn from a token's orthography.  
The features are processed by a Bi-LSTM, which captures the surrounding context of entities.
To generate a tag sequence a CRF in combination with a fully connected layer and final softmax classification is used.
The CRF determines the most likely sequence of tags for a feature sequence, enforcing tagging consistencies that are unknown to lower layers of the classifier. 
The Bi-LSTM-CRF was implemented in PyTorch v1.4.0~\cite{NEURIPS2019_9015} and the Word2Vec model with Gensim v3.8.1~\cite{rehurek_lrec}.
All models were trained with the standard configuration~\cite{schindler2020investigating} with early stopping based on the development set.
Without further optimization, a Precision of 0.85 (0.82), Recall of 0.81 (0.77), and Fscore of 0.83 (0.79) was achieved for test and development set.
An overview of the results is given in \tabref{tab:software_recog}.

To test the potential generalisation of models trained on \corpus{} to previously unseen full text articles from PMC, training was performed in a 5-fold cross validation on the annotated full texts. 
An FScore of .791 ($\pm .039$) was achieved with 4 training runs on each fold.
Given that the baseline result is within the standard deviation of the Fscore, this demonstrates the generalisability of the corpus and suggests it as training corpus for full text analysis on PMC.

\begin{table*}
    \caption{Summary of NER results. Mention type is only considered correct if the most specific sub-type is recognized.}\label{tab:software_recog}
    \begin{tabularx}{\textwidth}{X|*{1}{>{\raggedleft\arraybackslash}p{1.2cm}}|*{5}{>{\raggedleft\arraybackslash}p{.39cm}}|*{1}{>{\raggedleft\arraybackslash}p{.4cm}}|*{4}{>{\raggedleft\arraybackslash}p{.39cm}}|*{1}{>{\raggedleft\arraybackslash}p{.4cm}}|*{9}{>{\raggedleft\arraybackslash}p{.39cm}}|*{1}{>{\raggedleft\arraybackslash}p{.4cm}}}
        
        \toprule
        \multicolumn{2}{r}{}& \rot{Application} & \rot{Plugin} & \rot{OS} & \rot{PE} & \rot{Co-reference} & \rot{Software type} & \rot{Allusion} & \rot{Usage} & \rot{Creation} & \rot{Deposition} & \rot{Mention Type} & \rot{Abbreviation} & \rot{Developer} & \rot{Extension} & \rot{Alt. Name} & \rot{License} & \rot{Citation} & \rot{Release} & \rot{URL} & \rot{Version} & \rot{Additional Info}\\ 
        \midrule
        \multirow{4}{*}{\rot{Test}}&Precision&.79&.64&1.0&.98&.00&\textbf{.80}&.68&.77&.87&.76&\textbf{.77}&.91&.88&.60&.25&.91&.92&.88&.91&.92&\textbf{.89}\\
        & Recall&.84&.32&.70&1.0&.00&\textbf{.77}&.18&.84&.51&.57&\textbf{.72}&.59&.87&.60&.25&.71&.82&.78&.98&.92&\textbf{.87}\\
        & FScore&.81&.43&.82&.99&.00&\textbf{.78}&.29&.80&.64&.65&\textbf{.74}&.71&.88&.60&.25&.80&.87&.82&.95&.92&\textbf{.88} \\
        & Support &415&78&30&63&4&\textbf{590}&71&438&53&28&\textbf{590}&17&110&5&4&14&120&9&53&190&\textbf{522} \\
        \midrule
        \multirow{4}{*}{\rot{Devel}}&Precision&.73&.65&1.0&.95&.75&\textbf{.74}&.68&.78&.97&.79&\textbf{.79}&.89&.91&1.0&.25&1.0&.93&.67&.93&.94&\textbf{.89}\\
        & Recall &.80&.27&1.0&.95&.33&\textbf{.70}&.28&.80&.50&.53&\textbf{.67}&.47&.93&.60&.17&.80&.78&.86&.94&.98&\textbf{.85}\\
        & FScore &.76&.39&1.0&.95&.46&\textbf{.72}&.40&.79&.66&.63&\textbf{.71}&.62&.92&.75&.20&.89&.85&.75&.94&.96&\textbf{.87}\\
        & Support &303&113&19&55&9&\textbf{499}&67&336&60&36&\textbf{499}&17&87&5&6&10&89&7&70&139&\textbf{430}\\
    \end{tabularx}
\end{table*}

\subsubsection{Relation Extraction} RE is performed for all software-related entities within one sentence by using a Random Forest classifier with manually engineered input features.
For each pair of entities and its context, the following features were considered:
(1)~entity order,
(2)~entity types,
(3)~entity length in tokens and characters,
(4)~entity distance in tokens and characters,
(5)~sub-strings relation between entities, and
(6)~acronym relation between entities. 
Acronyms were generated by string normalization, exclusion of stop words, and removal of trailing numbers. 
RE is evaluated by Precision, Recall and FScore and summarized in \tabref{tab:add_info_recog}.
Overall, relations were extracted with a performance of .88 FScore.

\begin{table*}
    \caption{Summary of RE results for relating an entity to its base entity.}\label{tab:add_info_recog}    
    \begin{tabularx}{\textwidth}{*{1}{>{\raggedleft\arraybackslash}p{.15cm}}|*{1}{>{\raggedleft\arraybackslash}p{1.2cm}}|*{11}{>{\raggedleft\arraybackslash}p{0.9cm}}|*{10}{>{\raggedleft\arraybackslash}p{1.03cm}}}
        
        \toprule
        \multicolumn{2}{r}{}& \rot{Abbreviation} & \rot{Developer} & \rot{Extension} & \rot{AlternativeName} & \rot{License} & \rot{Citation} & \rot{Release} & \rot{URL} & \rot{Version} & \rot{Plugin} & \rot{Specification} & \rot{Relations}\\ 
        \midrule
        \multirow{4}{*}{\rot{Test}}&Precision&1.0&.94&1.0&1.0&.88&.87&.67&.92&.97&.76&1.0&\textbf{.92}\\
        & Recall&.94&.93&.80&1.0&.50&.84&.44&.87&.93&.56&.12&\textbf{.85}\\
        & FScore&.97&.94&.89&1.0&.64&.86&.53&.89&.95&.65&.22&\textbf{.88}\\
        & Support&17&111&5&4&14&121&9&53&190&39&8&\textbf{571}\\
        \midrule
        \multirow{4}{*}{\rot{Devel}}&Precision&1.0&.91&1.0&.83&1.0&.87&.75&.74&.96&.62&.75&\textbf{.87}\\
        & Recall&1.0&.99&1.0&.83&.40&.79&.86&.87&.96&.57&.50&\textbf{.87}\\
        & FScore&1.0&.95&1.0&.83&.57&.83&.80&.80&.96&.60&.60&\textbf{.87}\\
        & Support&17&87&5&6&10&90&7&70&139&35&6&\textbf{472}\\
    \end{tabularx}
\end{table*}

\subsubsection{Disambiguation} Software entities are disambiguated by hierarchical clustering. 
The method was advanced from the approach described in Schindler et al.~\cite{schindler2020investigating} by considering the available additional information.
The following features were computed for all pairs of entity mentions:
(1)~Levenshtein distance and sub-string relation between entity strings, associated URLs, and associated Developers,
(2)~known DBpedia alternative software names as a form of distant supervision, 
(3)~Levenshtein distance on normalized strings (casing, removal of stop words and trailing numbers), and
(4)~exact match comparison of automatically generated acronyms from normalized strings. 
The individual linking steps based on the feature matrices were also manually engineered and their details can be found in the published implementation, see \secref{sec:available}.
Entity disambiguation was evaluated based on all entity pairs by comparing predicted links between pairs against the gold standard. 

Considering all possible entity tuples, we achieve Precision 0.99, Recall 0.96 and Fscore 0.97. 
As this includes all mentions where the string is an exact match, the results were also calculated excluding exact matches: Precision 0.96, Recall 0.91, Fscore 0.93.
However, with exact string matches a false positive case was found for the software mention ``VBA'', once referring to Visual Basic for Applications (\url{https://www.wikidata.org/wiki/Q667566}) and three times to the Variational Bayesian Analysis toolbox (\url{https://mbb-team.github.io/VBA-toolbox/}).

\subsection{Software in Scholarly Articles}
\label{sec:soft_in_schol}

The recently published software citation principles~\cite{Katz2021} state that a software citation should be accompanied by the version, the developer and a URL.
By employing \corpus{} as a training set for information extraction, a large scale knowledge graph could be constructed to gain an understanding of software citation in science.
By querying the \corpus{} knowledge graph directly, we found that out of 3123 software mentions (application or plugin) in 1090 documents, only 1227 (39\%) are accompanied by a version, 733 (23\%) by a developer, 132 (4\%) by a URL, and 513 (16\%) by a formal citation. 
When considering version and developer at the same time, the number decreases to 520 (17\%) software mentions in 427 documents.
When looking at software depositions, out of 170 statements in 105 articles, 164 (96\%) are accompanied by a URL, 5 (3\%) by a version, and 49 (29\%) by a license. 
By exploiting the linked nature, \eg in combination with the MAKG~\cite{Faerber2019} or by applying supervised extraction pipelines trained on our corpus, further exploration of changes over time can be done. 

\section{Limitations}

\corpus{} was created through a careful manual annotation process of OA articles from PLoS and PubMed Central. 
While the IRR reflects very good agreement ($\forall\kappa{\approx}0.8$, $\forall\,$F1${>}0.72$), this does not imply that all entities of interest were discovered.
This is also true for disambiguation.
A challenging aspect for the annotation was distinguishing between applications and plugins. 
Similarly, the distinction between mentions of algorithms and software was hard, as authors tend to name software based on the implemented algorithm.
We found that external knowledge was required, usage of which is also suggested for automatic software recognition. 

Through its selection, see \secref{sec:article_selection}, \corpus{} is biased towards methods sections.
However, an analysis of PubMed fulltexts showed that 86\% of software mentions do appear in methods sections. 
Additionally, we were able to show that the corpus generalizes to unseen full text articles, see \secref{sec:NER}.

The frequencies for some types of entities are still low despite enrichment, for instance, with software creation statements.
We believe that this reflects the general rareness of those entities within articles of the represented domains.
Furthermore, the distribution of software and mention types might be a direct result from the selection of articles. 
The Softcite~\cite{https://doi.org/10.1002/asi.24454} dataset shows that software is potentially used less or cited less in economics. 

Finally, we observed few cases (${<}10$) where software and provided additional information stood in obvious conflict, \eg the provided developer does not reflect the actual developer.
We decided to still annotate the information to represent the author's intent to provide additional information for the software.

\section{Conclusion}
\corpus{} is a gold standard knowledge graph that covers software type, mention type and additional information, as well as their textual relations in scholarly articles.
It covers many concepts regarding software citations that have previously only been investigated individually or have not yet been explored, \eg different mention types and linking of entity names.
\corpus{} was created by manually annotating 1367 PMC OA articles covering a broad range of publications in life sciences and related disciplines.
Identified entities were linked to unique identifiers. 
The high quality of all manual annotation processes is underlined by very good agreement regarding textual annotations. 

We believe that \corpus{} is a valuable resource for the community as only few data for software mentions in scholarly texts is publicly available.
We sketched different use cases by performing analyses about software mentions contained in \corpus{} and their completeness.
We assume that the corpus will find most application as a gold standard for supervised machine learning and will contribute to establish an overview of the software infrastructure underlying the scientific progress.
To this end, baseline results for automatic information extraction were established. 
While baseline results are meant to demonstrate how the data may contribute to high-quality information extraction models, more sophisticated models will be able to improve the performance and enable further large-scale analysis of software usage in science.

\subsubsection*{Acknowledgements}
We thank Tazin Hossain and Hafiz Ghazi Sultan for their help with the annotation. 
This work was financially supported by the Deutsche Forschungsgemeinschaft (DFG, German Research Foundation) as part of the projects SFB 1270/2 (grant: 299150580) and ScienceLinker (grant: 404417453).

\bibliographystyle{ACM-Reference-Format}
\bibliography{softwarementions}
\end{document}